\documentclass[runningheads]{svmult}
\usepackage{makeidx}   % allows index generation
\usepackage{graphicx}  % standard LaTeX graphics tool
\usepackage{subeqnar}  % subnumbers individual equations
\usepackage{multicol}  % used for the two-column index
\usepackage{cropmark} % cropmarks for pages without
\usepackage{physprbb}  % centered layout of diverse elements, etc.
\newcommand{\greeksym}[1]{{\usefont{U}{psy}{m}{n}#1}}
\newcommand{\umu}{\mbox{\greeksym{m}}}

\newcommand{\farcm}{\hbox{$.\mkern-4mu^\prime$}}

\newcommand{\degr}{\hbox{$^\circ$}}

\begin{document}

\title*{Measuring Stellar and Dark Mass Fractions \protect\newline in Spiral Galaxies}

\toctitle{Measuring Stellar and Dark Mass Fractions \protect\newline in Spiral Galaxies}

\titlerunning{Stellar and Dark Mass Fractions in Spirals}

\author{Thilo Kranz
\and Adrianne Slyz
\and Hans-Walter Rix}

\authorrunning{Kranz et al.}

\institute{Max-Planck-Institut f\"ur Astronomie, K\"onigstuhl 17, 69117
Heidelberg, Germany}

\maketitle

\begin{abstract}
We explore the relative importance of the stellar mass density as compared to
the inner dark halo, for the observed gas kinematics thoughout the disks of
spiral galaxies. We perform hydrodynamical simulations of the
gas flow in a sequence of potentials with varying the stellar contribution to the
total potential. The stellar portion of the potential was derived empirically from
K-band photometry.
The output of the simulations -- namely the gas
density and the gas velocity field -- are then compared to the observed spiral
arm morphology and the H$\alpha$ gas kinematics.
We solve for the best matching
spiral pattern speed and draw conclusions on how massive the stellar disk can
be at most. For the case of the galaxy NGC 4254
(Messier 99) we demonstrate that the prominent spiral arms of the stellar component
would overpredict the non-circular gas motions unless an axisymmetric dark halo
component adds significantly in the radial range ${\rm R_{exp} < R < 3\, R_{exp}}$.
\end{abstract}

\section{Introduction}
In almost all galaxy formation scenarios, non-baryonic dark matter plays an important role.
Today's numerical simulations of cosmological structure evolution
quite successfully reproduce the observed galaxy distribution in the universe \cite{kau}.
While galaxies form and evolve inside dark
halos their physical appearance depends strongly on the local star formation and merging
history. At the same time the halos evolve and merge as well.
According to the simulations, we expect that the dark matter is important
in the inner parts of galaxies \cite{NFW1},\cite{NFW2} and that
it thus has a considerable influence on the kinematics. 
These predictions are in contrast to some studies which indicate that galactic stellar
disks - at least of barred spiral galaxies -  alone dominate the
kinematics of the inner regions \cite{deb}. Apparently this is also
the case in our own Milky Way \cite{ger}. \\
Determining individual mass fractions of the luminous and dark matter is not a
straightforward task. The rotation curve of a disk galaxy is only sensitive to the
total amount of gravitating matter, but does not allow the distinction between the two
mass density profiles. For a detailed analysis it is necessary to adopt more refined
methods to separate out the different profiles. Previous investigations used for example
knowledge of the kinematics of rotating bars \cite{wei} or the
geometry of gravitational lens systems \cite{mal}. 
Here we would like to exploit the fact, that the stellar mass in disk galaxies is often
organized in spiral arms, i.e.~in coherent non-axisymmetric structures. In most proposed
scenarios, the dark matter, however,
is collisionless and dominated
by random motions. Therefore it is not susceptible to spiral structures.
If the stellar mass dominates, the spiral arms, as traced by the near infrared (NIR) light,
should induce considerable non-circular motions in the gas, that manifest themselves
as velocity ''wiggles'' in observed gas kinematics. Using hydrodynamical gas
simulations we are able to predict these velocity wiggles
and compare them to the observations. Hence the contribution of the perturbative forces with
respect to the total forces can be determined quantitatively and
can be used to constrain the stellar disk to halo mass ratio. 

\section{Observations}

\begin{figure}[t]
\begin{center}
\includegraphics[width=.7\textwidth]{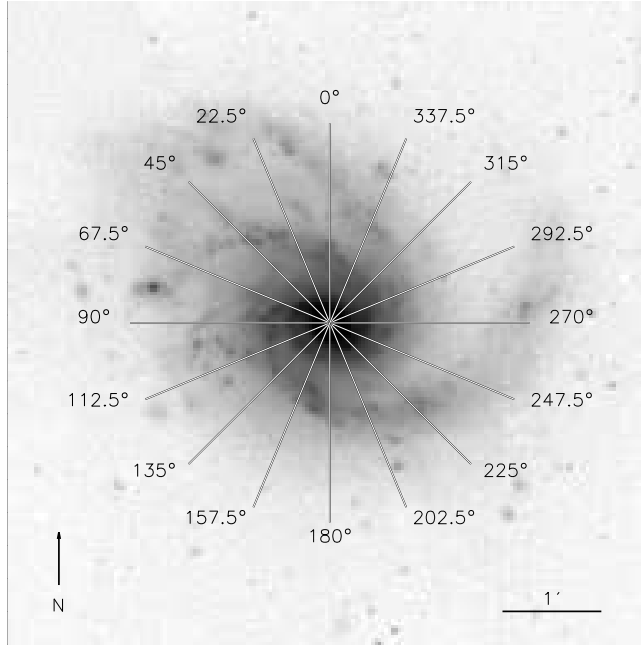}
\end{center}
\caption[]{Near infrared K'-band ($2.1\;$\umu m) image of NGC 4254 with a total exposure time
of 20 minutes at the Calar Alto
$3.5\;$m telescope. Bright foreground stars are masked out. The overlay shows the slit orientations
of the spectrograph. We took 8 longslit spectra (angles labeled in bold font) crossing the
galaxy's center to measure the 2D velocity field}
\label{n4254slits}
\end{figure}

For this analysis we need data to provide us with information on the stellar
mass distribution and on the gas
kinematics of a sample of galaxies. To map the stellar surface mass density
it is most desirable to take
NIR images of the galaxies, because in this waveband dust extinction and
population effects are minimized \cite{RR}.
During two observing runs in May 1999 and March 2000, we obtained photometric data
for ~20 close-by NGC galaxies. We
used the Omega Prime camera at the Calar Alto $3.5\;$m telescope with the K-band
filter (K' at $2.1\;$\umu m). It provides us with a field
of view of 6\farcm76 x 6\farcm76. Figure \ref{n4254slits} shows the K-band image of
the Messier galaxy M99. The labeled one arcminute scale bar translates to 5.8\, kpc
within the galaxy. \\
The kinematic data were obtained with the TWIN, a longslit spectrograph at the $3.5\;$m
telescope. For a
reasonable coverage of a galaxy's velocity field, we needed to take 8 slit
positions (or 16 position angles) across the entire disk of the
galaxy (also displayed in Figure \ref{n4254slits}). We chose longslit-spectroscopy
rather than fabry-perot interferometry because of its higher spectral resolution and
better sensitivity to faint emission. So far we were able to collect
complete sets of longslit spectra for only four
galaxies, mostly due to only moderate weather conditions during the spectroscopy runs.

\section{First results}
\begin{figure}[t]
\begin{center}
\includegraphics[width=.7\textwidth]{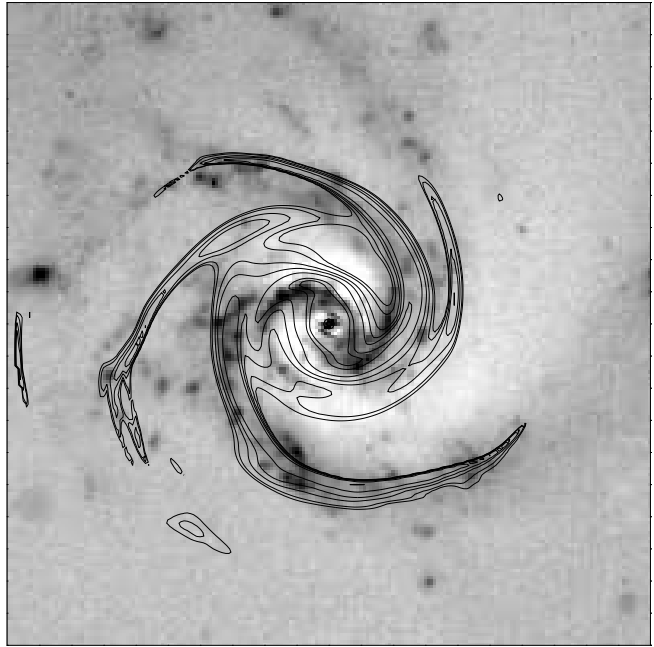}
\end{center}
\caption[]{Simulation results for the gas density. Here we depict the best fitting 
result (in contours) overlaid on a deprojected image of NGC\,4254. To enhance the
contrast of the spiral arms, an axisymmetric model of the galaxy has been subtracted
from the original image. We find an excellent match to the spiral morphology}
\label{gasdens}
\end{figure}
\begin{figure}[t]
\begin{center}
\includegraphics[width=.9\textwidth]{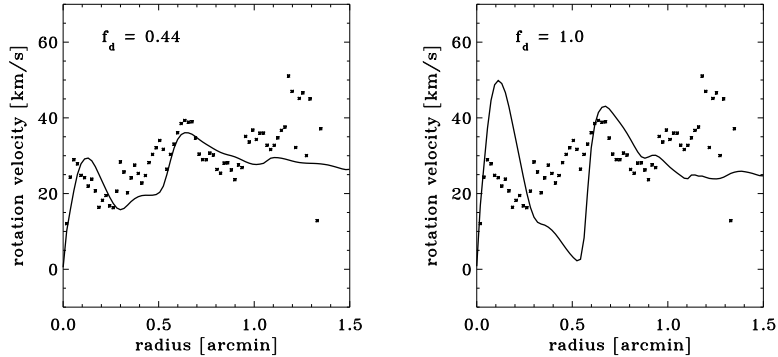}
\end{center}
\caption[]{Simulation results of the gas velocity. Here we compare two simulations
(continuous lines) with different disk mass fractions ${\rm f_d = M_{disk}/M_{maxdisk}}$
to the observed kinematics (data points). The maximal disk case (right) is clearly not
matching the observations well. Displayed is only one (135\degr) out of 16 slit position
angles}
\label{velcomp}
\end{figure}

As a pilot project, we analyzed the data of NGC 4254 (M99), a late type spiral galaxy
with strong arms, located in the Virgo cluster. Assuming a constant
stellar mass-to-light ratio in the K-band image, the
gravitational potential due to the stellar mass fraction was calculated by direct
integration over the whole light/mass
distribution. The mass-to-light ratio for the maximum disk
contribution was scaled by the
measured rotation curve. For the dark matter contribution we assumed an isothermal
halo with a core. To combine the
two components we chose a stellar mass fraction ${\rm f_d = M_{disk}/M_{maxdisk}}$
and added the halo with the variable
parameters adjusted to give a
best fit to the rotation curve. 
We used this potential as an input for the hydrodynamical gas simulations.\\ Figure
\ref{gasdens} presents the
resulting gas surface density, as it settles in the potential. The morphology of the
gas distribution is very sensitive to
the speed with which the spiral pattern of the galaxy rotates (pattern speed). In
figure \ref{gasdens} we printed the result of the
simulation whose spiral structure best matches the K-band image morphology. We
find quite good agreement for a pattern speed of
${\rm \Omega_{\rm p} = 23\; \rm km\, s^{-1}\, kpc^{-1}}$,
which places the corotation radius to 8.4\, kpc.\\ 
To quantify the relevance of the stellar mass component, we need now to fit the
velocities -- in particular the amplitude of the wiggles.
In figure \ref{velcomp} a comparison of the modelled and the measured rotation curves
are presented for the position
angle of 135\degr. The two panels show the rotation curves for different disk mass fractions:
less than half maximal and the maximum disk case.

\section{Discussion and outlook}
Although there is quite some scatter in the observed gas kinematics, we find that the
velocity jumps, which are apparent in the simulations for the maximum disk case are
too large to be in agreement with the measurements.
The inner part of the simulated rotation curve ($<$ 0\farcm3) is dominated by the
dynamics of the small bar, which
is present at the center of the galaxy. Its pattern speed might be different
from the one of the spiral's and
thus relate to a mismatch in the inner part of the rotation curve. 
We conclude that an axisymmetric dark halo is needed to explain the kinematics
of the stellar disk. The
influence of the stellar disk is submaximal in the sense that we don't find
strong enough velocity wiggles
in the observed kinematics as would be expected if the stellar disk was the
major gravitating source
inside the inner few disk scale lengths. 
How this conclusion might apply to other spiral galaxies will be the upcoming
issue of this project. We plan
to extend our analysis at first to the 3 other galaxies where we have already
now complete data sets.
Finally we intend to draw our final conclusions on a basis of a sample consisting
of 8 - 10 members. This
should be sufficient to determine reliable results about the luminous and dark
mass distributions in spiral
galaxies.

%INDEX%%%%%%%%%%%%%%%%%%%%%%%%%%%%%%%%%%%%%%%%%%%%%%%%%%%%%%%%%%%%%%%
% Please check with the editor of your book whether he plans to
% include a "mutual" subject index - if so, please code your entries
% in the standard syntax. For your own purposes you may print your
% "personal" index by using the following commands:
%
%\clearpage
%\addcontentsline{toc}{section}{Index}
%\flushbottom
%\printindex
%%%%%%%%%%%%%%%%%%%%%%%%%%%%%%%%%%%%%%%%%%%%%%%%%%%%%%%%%%%%%%%%%%%%%

\end{document}